\documentclass[sigconf]{acmart}
\usepackage{url}
\usepackage{graphicx}
\usepackage{algorithm}
\usepackage[noend]{algpseudocode}
\usepackage{amsfonts}
\algrenewcommand\algorithmicrequire{\textbf{Input:}}
\algrenewcommand\algorithmicensure{\textbf{Output:}}
\usepackage{balance}
\usepackage{color}
\usepackage{dsfont}
\usepackage{tabularx}
\usepackage{bm}
\usepackage{array}
\usepackage{booktabs}
\usepackage[inline]{enumitem}
\usepackage{subcaption}
\usepackage{multirow}
\newcolumntype{P}[1]{>{\centering\arraybackslash}p{#1}}
\newcolumntype{M}[1]{>{\centering\arraybackslash}m{#1}}

\DeclareMathOperator*{\argmin}{arg\,min}
\newcommand{\Break}{\State \textbf{break} }
\newcommand{\para}[1]{\smallskip\noindent\textit{#1.}}
\newtheorem{theo}{Theorem}
\acmConference[IWCTS '21]{14th International Workshop on Computational Transportation Science (IWCTS '21)}{November 1, 2021}{Beijing, China}
\acmBooktitle{14th International Workshop on Computational Transportation Science (IWCTS '21),
  November 1, 2021, Beijing, China}
\acmPrice{15.00}
\acmISBN{978-1-4503-9999-9/18/06}

\begin{document}

\title{Collective Shortest Paths for Minimizing Congestion on Temporal Load-Aware Road Networks}

\author{Chris Conlan}
\affiliation{
  \institution{University of Warwick}
  \city{Coventry}
  \country{United Kingdom}
  \postcode{CV4 7AL}
}
\email{chris.conlan@warwick.ac.uk}

\author{Teddy Cunningham}
\affiliation{%
  \institution{University of Warwick}
  \city{Coventry}
  \country{United Kingdom}
  \postcode{CV4 7AL}
}
\email{teddy.cunningham@warwick.ac.uk}

\author{Gunduz Vehbi Demirci}
\affiliation{%
  \institution{University of Warwick}
  \city{Coventry}
  \country{United Kingdom}
  \postcode{CV4 7AL}
}
\email{gunduz.demirci@warwick.ac.uk}

\author{Hakan Ferhatosmanoglu}
\affiliation{%
  \institution{University of Warwick}
  \city{Coventry}
  \country{United Kingdom}
  \postcode{CV4 7AL}
}
\email{hakan.f@warwick.ac.uk}

\renewcommand{\shortauthors}{Conlan and Cunningham, et al.}

\begin{abstract}
Shortest path queries over graphs are usually considered as isolated tasks, where the goal is to return the shortest path for each individual query. 
In practice, however, such queries are typically part of a system (e.g., a road network) and their execution dynamically affects other queries and network parameters, such as the loads on edges, which in turn affects the shortest paths. 
We study the problem of collectively processing shortest path queries, where the objective is to optimize a collective objective, such as minimizing the overall cost. 
We define a temporal load-aware network that dynamically tracks expected loads while satisfying the desirable `first in, first out' property.  
We develop temporal load-aware extensions of widely used shortest path algorithms, and a scalable collective routing solution that seeks to reduce system-wide congestion through dynamic path reassignment.
Experiments illustrate that our collective approach to this NP-hard problem achieves improvements in a variety of performance measures, such as, i) reducing average travel times by up to 63\%, ii) producing fairer suggestions across queries, and iii) distributing load across up to 97\% of a city's road network capacity. 
The proposed approach is generalizable, which allows it to be adapted for other concurrent query processing tasks over networks.
\end{abstract}

\copyrightyear{2021}
\acmYear{2021}
\setcopyright{acmcopyright}\acmConference[IWCTS'21]{14th ACM
SIGSPATIAL International Workshop on Computational Transportation Science
}{November 1, 2021}{Beijing, China}
\acmBooktitle{14th ACM SIGSPATIAL International Workshop on Computational
Transportation Science (IWCTS'21), November 1, 2021, Beijing, China}
\acmPrice{15.00}
\acmDOI{10.1145/3486629.3490691}
\acmISBN{978-1-4503-9117-7/21/11}



\maketitle

\section{Introduction}
\label{s:intro}
The shortest path query over networks is a widely studied problem with well-known and efficient solutions.
The most common setting is to consider each query in isolation, where the objective is to optimize each individual path. 
However, in practice, multiple queries are executed simultaneously and each query influences the cost incurred by other queries. 
Navigational services are a real-world example of executing concurrent shortest path queries.
Simply targeting the fastest route\footnote{We use the terms shortest paths and fastest paths/routes interchangeably} for each query independently can induce congestion as the recommended routes are likely to share common edges. 
Consequently, one must aim for an aggregate optimization goal that seeks alternative routes for some vehicles in order to reduce system-wide congestion and decrease total travel times.  
To formally address the problem, in this paper, we introduce a temporal load-aware setting in which the network tracks expected future load in the system based on the queries' origin and destinations.
An arrival time function uses the load to penalize congestion, which is integrated into the shortest path algorithms so that congestion is proactively avoided.
Our goal is to address the problem in a collective manner (i.e., we seek optimal routes for a set of queries concurrently), rather than calculating routes independently for chronologically encountered queries.

\begin{figure}[t]
    \centering
    \includegraphics[width=0.98\columnwidth]{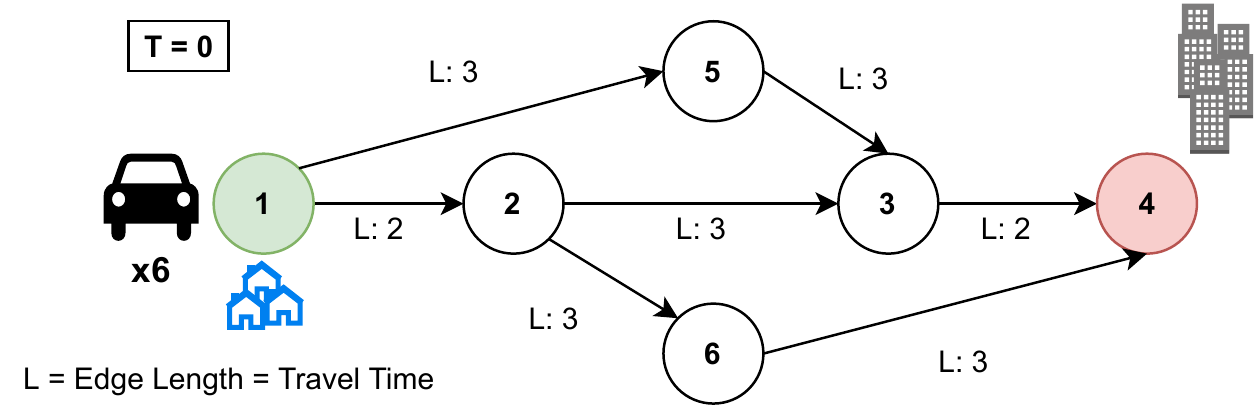}
    \caption{Motivating Example}
    \label{fig:motivating-example}
\end{figure}

Figure \ref{fig:motivating-example} is a simple illustration of the principle of collectively processing shortest path queries, and demonstrates how anticipating future congestion can provide alternative paths that are globally optimal. 
In this example, there is a demand to route six vehicles from node 1 to node 4 and each edge has capacity for four cars to travel at maximum speed.
If loads exceed capacity, we assume all vehicles experience delays.
A traditional navigational service may route all six vehicles along path 1-2-3-4, which would cause congestion and delays along these edges. 
A collective solution would direct two vehicles along route 1-5-3-4, two vehicles along 1-2-3-4, and two vehicles along route 1-2-6-4.
Although some travelers experience an additional time cost (equal to one unit) compared to the uncongested shortest path, these routes result in less congestion and delay for individuals and the system. 
We note that we consider multiple origins and destinations in the problem we study.

The problem of collectively processing shortest path queries (CP-SPQ) is formalized as minimizing the aggregate cost for concurrent queries in a network that is both time- \emph{and} load-dependent. 
We develop a collective solution that enables proactive decisions to be made, based on the expected future loads, by adjusting edge values, and dynamically selecting the paths with the earliest expected arrival at their destination. 
To do so, we first define the notion of temporal load-aware networks (TLAN), which maintain expected future load along each edge within a given time interval, with a proof to satisfy the desirable first-in, first-out (FIFO) property. 
This property enables a TLAN to guarantee viable acyclic routes exists and that waiting on nodes cannot result in shorter routes.
We utilize a context-aware arrival time function to penalize edges that are expected to be congested, and this controls transitions within the TLAN whilst satisfying FIFO requirements. 

We then develop temporal load-aware adaptations of the widely used A* and top-$k$ algorithms. 
These adaptations detect expected future congestion in the network, and dynamically provide alternative routes to avoid congestion by enabling queries, past and present, to interact via the TLAN.
We finally propose a scalable solution for collectively and concurrently computing shortest paths that aims to optimize system-level travel times. 
As opposed to chronological processing of queries, our approach aims to find a more advantageous order for processing queries by iteratively finding paths that returns the lowest arrival time. 
The processing is embarrassingly parallel thus ensuring scalability. 
We discuss how our work can be integrated into navigational services, enabling them to benefit from collective processing. 
We enhance the efficiency of the collective solution by utilizing machine learning to restrict the search space of possible routing options. 
We maintain accuracy by continuously tracking computations between iterations to reduce redundancies.

Experiments on real-world data show that our collective approach achieves significant improvements in congestion reduction, utilization of the network's capacity, and distributing the load across a more diverse set of edges. 
For example, the proposed approach achieves time savings (compared to existing solutions) of up to 4.8 minutes per trip in Porto (17\% saving), and 3.4 minutes in New York (20\% saving); across all users this equates to saving thousands of hours for the city. 
The results are consistent when the collective system incorporates only a portion of the load (e.g., when an operator does not observe the entire traffic flow, or users do not follow recommended paths). 
Our solution is able to reach up to 97\% load distribution, utilizing almost all capacity on the roads, which represents a 30\% improvement on baseline algorithms.

Our contributions can be summarized as follows:
\begin{itemize}
    \item We study the CP-SPQ problem with a novel formalization for temporal-load aware networks and a context-aware arrival time function, which enables the desirable FIFO property.
    \item We propose temporal load-aware adaptations of popular algorithms and a scalable solution that collectively processes shortest path queries with significant reductions in costs.
    \item We provide a framework for studying the CP-SPQ problem in large TLANs, including a set of performance metrics, covering total costs, network utilization, and fairness. 

\end{itemize}

\section{Related Work}
\label{s:related-work}

A wide variety of shortest path problems have been studied extensively with many efficient solutions \cite[see][]{madkour2017survey,zhan1998shortest}.
A particularly useful setting is to model the network with temporal data that brings time savings for the average journey \cite{m24}. For example, \citet{m2} maintain a temporal network with a static structure and time-varying features (e.g., travel time), which improves space and algorithmic efficiency.
In our work, the network also incorporates an arrival time function to capture time and load variations along edges, as well as the current and future state of the network.

Route diversification is a common approach for congestion alleviation in road networks~\cite{cheng2019shortest}. 
\citet{2020-sub-33} present a routing algorithm for a network of autonomous vehicles, by generating a set of viable alternative routes and then assigning one for each query. 
This approach considers each individual vehicle (rather than as a collective optimization problem) and assigns the routes randomly. 
An alternative approach is to adapt Yen's top-$k$ shortest paths algorithm \cite{m17} and select paths with high spatial diversity and low cost diversity \cite{cheng2019shortest}.
This approach seeks to optimize a single query but does not address the collective routing of multiple queries for different origins/destinations. 
\citet{nguyen2015randomized} adapt the A* algorithm to introduce diversity in route allocation. 
Their approach is to randomly perturb the edge costs to change the returned shortest path, but with minimal impact to the cost of the paths returned. 
\citet{jeong2015saint} propose a navigational system that necessitates physical infrastructure (e.g., through installation of ``base stations'') and with which seeks the path with ``shortest'' congestion contribution within an acceptable delay to the user. 
We propose and demonstrate the effectiveness of a collective approach, something that these works do not particularly address.

The principle of a user equilibrium \cite{wardrop1952road} on a road network motivates other works.
That is, if each user takes their best path, an equilibrium will be reached meaning no driver can improve their travel time by taking an alternative route. 
In practice, as the road network is dynamic and information on the state of the network is imperfect, this equilibrium is never reached. 
Dynamic user equilibrium approaches \cite{pan2012proactive, el2017traffic} route and then re-route vehicles dynamically, responding in real-time to road conditions. 
These approaches, however, are not inherently collective as they optimize for each user rather than the overall system. 
Our approach also proactively avoids congestion by integrating temporal load-aware features of the network into the routing and shortest path computation.

The broad notion of system optimal routing has been explored in various branches of literature, such as the studies on ``socially optimal'' routing~\cite{2020-sub-31} and 
linear programming based optimization~\cite{angelelli2016system}. 
\citet{2020-sub-28} use an iterative approach of routing and rerouting trajectories until an acceptable stopping criteria is reached. 
Our work is fundamentally different in that we collectively route vehicles based on the expected workload within the system in order to proactively avoid congestion.
Their system also depends on two-way connectivity between the system and the users so that real-time traffic conditions are detected.
\citet{motallebi2020streaming} propose a solution that avoids intersecting routes and, to do so, they maintain a heatmap of normalized travel times over the network (i.e., ``reservation'' graph that tracks expected load and learned temporal patterns from historic data). 
However, shortest path computations do not dynamically account for these features; instead they calculate the lowest impact route from a candidate set.
The approach also handles queries chronologically, as opposed to collectively. 
\section{Temporal Load-Aware Setting}
\label{s:prob-set-up}

In this section, we introduce the temporal load-aware setting, and formally define the problem we study.
The challenging task of formulating an environment that is both time- and load-dependent is essential in enabling our solutions to respond dynamically to the current and future state of the network. 
It is desirable that this dynamic environment is FIFO-compliant so that shortest path algorithms can find a realistic solution. 
We aim to achieve efficiency with a dynamic arrival time function that operates on a static network structure, which means that we do not need to maintain multiple impressions of the network and its features.

\subsection{Problem Definition}
\label{ss:preliminaries}
To model a road network, we define a directed graph $\mathcal{G}(\mathcal{V},\mathcal{E})$ with vertex set $\mathcal{V}$ representing intersections, and an attributed edge set $\mathcal{E}$ representing roads. 
Edge attributes include vehicle capacity~$F_{ij}$, edge length $\delta_{ij}$, speed limit $z_{ij}$, and minimum travel time $\Upsilon_{ij}$.
We use the $E = |\mathcal{E}|$ notation to denote set cardinality.

A hybrid discrete-continuous time domain is utilized to reduce the space complexity of the problem, and to introduce time intervals to better record load on edges (where more specific time predictions could not be guaranteed).
Time intervals are given by $\mathcal{T} = \{\tau_1, \tau_2, \ldots, \tau_n, \ldots, \tau_T\}$, where the length of an interval is $\tau_{n+1} - \tau_n = I$.
Edge loads are recorded for each discrete time interval, and the arrival time at a node is given in the continuous time domain, denoted as $t$. 
We define the load as the number of vehicles who occupy an edge at any point during the time interval.

The edge-load matrix~(ELM), given by $\mathcal{L}(\mathcal{E},\mathcal{T})$, contains the load for each edge $e_{ij} \in \mathcal{E}$, for each (future) time interval $\tau \in \mathcal{T}$. 
We use $l(e_{ij}, \tau)$ to denote the load on edge $e_{ij}$ during $\tau$. 
The ELM tracks the expected temporal variation in load across the network.  
It is also key in defining a temporal load-aware network -- a fundamental part of the solution to the CP-SPQ problem.

\begin{definition}[Temporal Load-Aware Network]
A network given by $\mathcal{G}(\mathcal{V},\mathcal{E})$ in which the load along each edge $e_{ij} \in \mathcal{E}$ is tracked with respect to each discrete time interval in $\mathcal{T}$ under the edge-load matrix $\mathcal{L}(\mathcal{E},\mathcal{T})$. 
\end{definition}

The query set $\mathcal{Q}$ contains shortest path queries $q_{sdt} \in \mathcal{Q}$ from $v_s$ to $v_d$, starting at time $t$. 
For each query $q_{sdt}$, we determine a path $p_{sdt}$, which is assigned to the path set $\mathcal{P}$.  
Each path $p_{sdt}$ consists of a sequence of edge-time interval pairs $(e, \tau)$ to connect $v_s$ with $v_d$.
Each query has a theoretical shortest path $\phi_{sdt}$ under free-flow conditions. 
We use $|\cdot|$ notation to denote the length of a path (synonymous with travel time).
The assigned path may have a greater cost than $\phi_{sdt}$, either due to congestion along the shortest path, or because an alternative path is selected as $\phi_{sdt}$ is too congested. 
Hence, the congestion penalty is:
\begin{equation}
    \pi_{sdt} = |p_{sdt}| - |\phi_{sdt}|
\end{equation} 

An edge-level arrival time function, $f_{ij}$, controls transitions in the network. 
It is a user-specified, non-negative function of edge attributes and the current load.
We denote the arrival time at $v_i$ as $a_i$. 
Hence, $a_j = f_{ij}(a_i)$, and the arrival time of $q_{sdt}$ at $v_d$ is $a_{sdt}$.

Our aim is to compute a path $p_{sdt}$ for each $q_{sdt}$, such that the total travel time of all $p_{sdt} \in \mathcal{P}$ is minimized, formally defined as:
\begin{equation}
    \textstyle
    \label{eq:objective-function}
    \text{min} \quad \sum^{Q}_{k=1} a^k_{d}
\end{equation}
\begin{equation}
    \textstyle
    \label{eq:demand}
    \sum\limits_{j:e_{ij}\in \mathcal{E}} x^k_{ij}(\tau) - \sum\limits_{j:e_{ji}\in \mathcal{E}} x^k_{ji}(\tau) = b^k_i
\end{equation}
\begin{equation}
    \textstyle
    \label{eq:source-dest}
    b^k_i = 
        \begin{cases}
            1  & \text{if } i=v_s^k\\
            -1 & \text{if } i=v_d^k\\
            0  & \text{otherwise}
        \end{cases}
\end{equation}
\begin{equation}
    \textstyle
    \label{eq:arrival-time-constraint}
    x^k_{ij}(\tau) \cdot f_{ij}(a^k_i) = x^k_{ij}(\tau) \cdot a^k_j
\end{equation}
\begin{equation}
    \textstyle
    \label{eq:at-node}
    x^k_{ij}(\tau) = 
        \begin{cases}
            1  & \text{if } a^k_i \leq \tau \leq a^k_j \\
            0  & \text{otherwise}
        \end{cases}
\end{equation}
\begin{equation}
    \textstyle
    \label{eq:capacity}
    l(e_{ij}, \tau) = \sum_{k=1}^{Q} x^k_{ij}(\tau)
\end{equation}
\noindent
where $x^k_{ij}(\tau)$ is a binary variable denoting whether $e_{ij}$ is traversed by a vehicle $k$ during $\tau$. 
\eqref{eq:demand} and \eqref{eq:source-dest} ensure that a valid path from $v_s^k$ to $v_d^k$ is established.
\eqref{eq:arrival-time-constraint} ensures that, if an edge is traversed (i.e., $x^k_{ij}(\tau) = 1$), the arrival times satisfy the requirements of the arrival time function: $f_{ij}(a^k_i) = a^k_j$.
\eqref{eq:at-node} and \eqref{eq:capacity} impose load constraints on edges. 
That is, if $e_{ij}$ is traversed during $\tau$, then $x_{ij}(\tau)\!=\!1$.
The total load $l(e_{ij}, \tau)$ is the number of vehicles using edge $e_{ij}$ in $\tau$, which is the sum of $x^k_{ij}(\tau)\!=\!1$ over all $Q$ vehicles.

The problem without the temporal load-aware constraint is NP-hard \cite{jahn2005system}, which motivates our heuristic-based approach.

\subsection{Arrival Time Function}
\label{ss:arrival-time-function}
A time- and load-dependent arrival time function dynamically computes travel times in TLANs, enabling proactive decisions about which edges to traverse. By using such an arrival time function we can maintain a static network structure, but vary weight edges dynamically which is space-efficient. 
Importantly, our solutions are not dependent on this specific arrival time function; any FIFO-compliant arrival time function is permissible.

\para{Flow Regimes}
When $l(e_{ij}, \tau) \leq F_{ij}$, an edge is in a \textit{free-flow regime}. 
In a road network, this means all vehicles are traversing the edge at (or below) the speed limit.
When $l(e_{ij},\tau) > F_{ij}$, an edge is in a \textit{congested flow regime}. 
This means that any vehicles that subsequently join this edge will experience slower travel times (thus ensuring safe distances between vehicles are maintained). 
Other methods for modeling the behavior of vehicles along edges would be appropriate as long as they are FIFO-compliant.

\para{Delay Exponent}
To control the rate at which speeds decrease along an edge, we introduce a load-dependent exponent, $\epsilon$, that ensures travel times are penalized when edges are congested. 
Naturally, the more congestion that exists on an edge the greater the delay. 
In free-flow, $\epsilon = 1$; in congested flow, $\epsilon$ decreases as load increases:
\begin{equation}
    \textstyle
    \label{eq:delay-exponent}
    \epsilon(\tau) =
    \begin{cases} 
       1                              & l(e_{ij},\tau) \leq F_{ij} \\
       \frac{1}{l(e_{ij},\tau) - F_{ij}}  & l(e_{ij},\tau) > F_{ij} 
    \end{cases}
\end{equation}

\para{Arrival Time Function}
The arrival time function for an edge $e_{ij}$ is:
\begin{equation}
    \textstyle
    \label{eq:arrival-time}
    f_{ij}(a_i)  = \tau_i +  \left(a_i-\tau_i\right)^{\epsilon(\tau_i)} + \Upsilon_{ij} = a_j
\end{equation}
where, $a_i$ and $a_j$ are the arrival times in the continuous domain at vertices $v_i$ and $v_j$ respectively, and $\tau_i = \lfloor a_i \rfloor$ (i.e., the discrete time interval in which $a_i$ lies).

In free-flow, the function returns the time in the continuous domain that a user arrives at $v_i$, plus the minimum travel time along an edge. 
As the load increases, the delay term tends to $I$. 
This bounds the delay term such that a user cannot wait until the following time interval to return an earlier arrival time -- an important characteristic that is crucial to ensuring FIFO compliance.

Figure \ref{fig:exponent_affect} shows how $a_j$ varies under different conditions.
As the edge load increases, the delay term increases and the time taken to traverse the edge increases non-linearly.  
As the load increases by a large amount, the effects of congestion on $a_j$ start to plateau (e.g., traffic is likely to be traveling almost as slow as possible). 

\subsection{First-In, First-Out (FIFO) Compliance}
\label{ss:fifo-proof}
FIFO networks have been integral for many routing tasks \cite{2020-sub-08,2020-sub-09,2020-sub-10,2020-sub-11}.  
Here, we demonstrate how our arrival time function for a TLAN is FIFO-compliant.
FIFO compliance guarantees that: 
\begin{enumerate*}[label={\alph*)}]
\item an acyclic shortest path exists in a TLAN, which guarantees a viable path exists for any $(s,d,t)$ combination; and
\item the sub-path of a shortest path is also a shortest path, which is necessary to guarantee results using Dijkstra's shortest path algorithm.
\end{enumerate*}
Our formulation guarantees that all vehicles exit edges in the same order in which they enter (i.e., time gains along an edge cannot be attained by delaying the departure from a particular node).
\begin{theo}
    The arrival time function, $f_{ij}(a_i)$, is FIFO-compliant.
\end{theo}
\begin{proof}
To show that $f_{ij}(a_i)$ is FIFO-compliant and waiting at a node cannot result in a quicker traversal time, we consider the two possible `wait' cases.

\para{Case 1 -- Waiting within a time interval}
If a user waits within a time interval until the next time interval, the only element of the arrival function that varies is the inner component of the delay term. This is a function of $a_i$, which increases as $a_i$ increases. Hence, the arrival time can only increase as $a_i$ increases within a time interval.

\para{Case 2 -- Waiting for the following time interval}
If a vehicle waits at a node until the following time interval, then the delay exponent can vary. We need to examine whether it is possible for a vehicle arriving late within a congested time interval to wait until the following time interval (with less congestion) and be given an earlier arrival time.
In this case, we are required to show that:
\begin{equation}
    \label{eq:fifo-case-2}
    f_{ij}(a_1) \leq f_{ij}(a_2) \qquad \forall \; a_1 \in \tau_{1}, \; a_2 \in \tau_{2}
\end{equation}
where $a_1$ and $a_2$ are two different arrival times at $v_i$ and $\tau_1 < \tau_2$.  The limiting case is the one in which there is maximum congestion during $\tau_1$, and no congestion during $\tau_2$.  For $\tau_1$, as the exponent tends to 0, $f_{ij}(a_1) \to  \tau_1 + I + \Upsilon_{ij}$.  And, for $a_2$, when there is no congestion, the delay exponent equals 1 and so, $f_{ij}(a_2) = \tau_2 + \Upsilon_{ij}$. Knowing that $\tau_2 = \tau_1 + I$, we can show that $\max(f_{ij}(a_1)) \leq \min(f_{ij}(a_2))$.
\end{proof}

\begin{figure}[t]
    \centering
    \includegraphics[width=0.9\columnwidth]{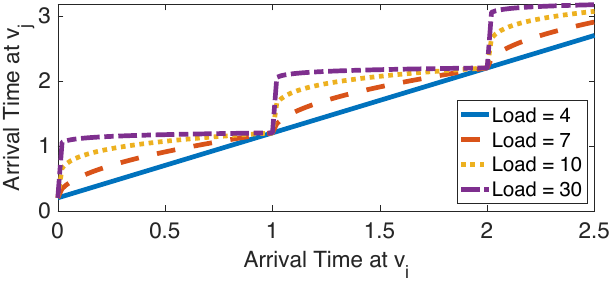}
    \caption{Effect on arrival time given varying loads along $\bm{e_{ij}}$, with $\bm{F_{ij}}$ = 5, and $\bm{\Upsilon_{ij} = 0.3\tau}$; arrival times given in terms of $\bm{\tau}$}
    \label{fig:exponent_affect}
\end{figure}

Figure \ref{fig:exponent_affect} also illustrates how $f_{ij}$ ensures that waiting until the next timestep can not allow an earlier arrival time.  
Any vehicle leaving in the first timestep will always arrive earlier than vehicles traveling along an uncongested edge in the second timestep.

\section{Collectively Processing Shortest Paths Queries}
\label{s:algorithms}

In this section, we present our approach to address the CP-SPQ problem. 
Initially we address the shortest path query in a TLAN. 
The context-aware arrival time function enables the algorithms to dynamically account for the expected future load in the system, thus proactively avoiding congestion.
The ordering of queries, however, is important as chronological ordering (or any other ordering) does not guarantee optimal results.
To address this issue, we then present CS-MAT, our solution for collectively processing queries.
CS-MAT searches for a path with the earliest arrival time from unallocated queries whilst avoiding re-computation of routes on-the-fly from a candidate set. 
Concurrently processing a large number of shortest path queries is a significant technical challenge, and so we demonstrate how our approach can be parallelized easily, which enables it to be applied to large networks and query workloads. 

\subsection{Shortest Path Query in a TLAN}
In this section, we present algorithms that solve the shortest path query in TLANs. 
By operating in a TLAN and updating the ELM after each query is processed, these solutions are able to detect expected future load within the system. Congested edges are therefore proactively avoided as the arrival time function penalizes these edges.
First, we present temporal load-aware A* (TLAA*) -- an adaptation of the classic A* algorithm \cite{hart1968formal}, and then we present a high-level description of Temporal Load-Aware Top-\textit{k} (TLAT$k$).

\subsubsection{Temporal Load-Aware A*}
\label{ss:lad-a*}
In TLAA* (Algorithm \ref{alg:load-aware-a-star}), new shortest paths are iteratively discovered and recorded with their actual cost (taking expected congestion into account), plus the expected cost to the destination node. 
At the next iteration, TLAA* visits the node with the lowest expected cost to the destination node.
The length of the shortest free-flow path is used as the expected cost function. 
This is kept in a pre-computed pairwise matrix. 
As the network is FIFO-compliant, TLAA* guarantees that the shortest path for any query and any load distribution can be found.

\para{Algorithm Outline}
First, we initialize three tracking vectors: $B$, $M$, and $H$. 
$B$ tracks the length of the best path to a given node in $\mathcal{G}(\mathcal{V},\mathcal{E})$ and we set $B(v)$ to $\infty$ initially.
$M$ tracks the predecessor node along the currently known shortest paths and we initialize $M(v)$ as $\emptyset$.
$H$ gives the expected cost from a given node to $v_d$ (Lines 1-2) and we initially set $H(v)\!=\!\infty$ for all $v\!\in\!\mathcal{V}$, except $v_s$ for which we set $H(v_s)\!=\!\phi_{sdt}$ (Line 2).
We first identify the shortest path in $H$ (Line 5); in the first iteration, this will be the source node. 
If the identified node is $v_d$, the algorithm terminates (Lines 6-7).  
Otherwise, we remove the current node from $\mathcal{V}'$ and obtain the time it arrives at $\bar{v}$, denoted as $\bar{a}$ (Lines 9-10), given by query start time plus the length of the best known path to the current node.  
The load-aware arrival times at neighboring nodes to $\bar{v}$ are then computed ($\mathcal{N}_{\mathcal{G}}(\bar{v})$ denotes the neighborhood set). 
If a shorter path to any node is found, $B$, $M$, and $H$ are updated (Lines 11-14).

\begin{algorithm}[t]
\small
\algrenewcommand\algorithmicindent{0.75em}%
	\caption{Temporal Load Aware A*}
	\label{alg:load-aware-a-star}
	\begin{algorithmic}[1]
	\Require{$\mathcal{G}(\mathcal{V},\mathcal{E})$, $\mathcal{L}(\mathcal{E},\tau)$, $q_{sdt}$}
	\Ensure{$p_{sdt}$}

    \State Initialize $B, M, H$
    \State $H(s) \gets \phi_{sdt}$
    \State $\mathcal{V}' \gets \mathcal{V}$
    \While{$\mathcal{V}' \neq \emptyset$} 
        \State $\bar{v} \gets \argmin_v (H)$ \Comment{Shortest path in H}
        \If{$\bar{v} = d$}
	        \State \textbf{return} $p_{sdt}$
	    \Else
            \State $\mathcal{V}' \gets \mathcal{V}' \setminus \bar{v}$ 
            \State $\bar{a} \gets q_{t} + B(\bar{v})$ \Comment{Time arriving at current node} 
            \ForAll{$v \in \mathcal{N}_{\mathcal{G}}(\bar{v})$} \Comment{Visit neighboring nodes}
                \State $a_j \gets f_{ij}(\bar{a})$
                    \If{$a_j < B(\bar{v})$} \Comment{Quicker path found}
                        \State Update $B, M, H$
                    \EndIf
            \EndFor
        \EndIf
    \EndWhile
	\end{algorithmic}
\end{algorithm}

\para{Running Time Efficiency}
The worst-case running time for TLAA* is $\mathcal{O}(V\log V + ET)$. 
$V\log V$ gives the number of searches in the graph for a new shortest path when common speed-up techniques are used (e.g., min-heaps, adjacency lists \cite{zhang2012improved}).
Each search necessitates a query on the ELM to return the expected future load.
The search on this matrix is bounded by $T$ (i.e., the maximum number of time intervals) and, as no edge will be traversed more than once, this yields $ET$.
Since the search is guided towards the target with an expected cost function, the running time is more efficient in practice.
The memory required to implement TLAA* is concentrated on the ELM, and is upper-bounded at $\mathcal{O}(ET)$. 
In practice, we find that operations on this matrix are executed in a matter of milliseconds (using our computational set-up described in Section \ref{ss:expt-set-up}).

\subsubsection{Temporal Load-Aware Top-\textit{k}}
\label{ss:temp-top-k}
TLAT$k$ chooses the optimal path among each query's pre-computed top-$k$ shortest paths given the known and expected load distribution.
By pre-computing possible paths we speed up online processing, but lose some depth in the search compared to TLAA*. In a low congestion system where users are likely to take their free-flow best path anyway, this approach may yield useful results. 
As a pre-processing step, we compute the $k$-shortest paths under free-flow conditions (e.g., by using Yen's algorithm \cite{m17}) and store them in a matrix.
As load builds up in the system, and some edges become congested, the fastest free-flow path will no longer necessarily be the best path to assign to the user. 
Within the temporal load-aware setting, we can calculate the expected cost of the $k$-shortest paths given the current (and expected) load, and assign to the user the fastest path. 
The running time complexity of this algorithm is $\mathcal{O}(kV)$. 

\subsection{Collective Search For Minimal Arrival Time (CS-MAT)}
\label{ss:cs-mat}
TLAA* guarantees the shortest paths in a TLAN for any \textit{individual} query. 
But, as the ELM is updated after processing each query, the load distribution can vary significantly depending on the order in which queries are processed.
To address this issue, we propose CS-MAT, a collective algorithm that iteratively finds the lowest possible arrival time query from a candidate set.

From within a candidate set of queries $\mathcal{Q}' \subseteq \mathcal{Q}$, CS-MAT uses any temporal load-aware algorithm (e.g., TLAA*) to find the path with the earliest arrival time at its destination node and assign this path. 
In most cases, this is not the query that is encountered chronologically, which often leads to the same query being `re-processed' multiple times. 
Once a query has been assigned, it is added to the ELM ensuring future queries can dynamically take account of its load impact on the system.

A na\"{i}ve way to perform these steps would be to simply re-process all unallocated queries after a query has been assigned a path and the ELM has been updated. 
 
This is inefficient and leads to many redundant calculations. 
Hence, CS-MAT restricts the search in $\mathcal{Q}$ for $\mathcal{Q}'$. 

\para{Predicting Congestion Penalty}
To restrict the search in $\mathcal{Q}$ for candidate queries, queries whose departure time is after the predicted arrival time of the current query are ignored, as they are unlikely to give an earlier arrival time. 
We can utilize a machine learning model, trained on representative queries, to predict a query's congestion penalty, $\pi^{pred}$.
The training query set is processed using TLAA*, and the congestion penalties are captured.
Using a feed-forward neural network, we convert $v_s$ and $v_d$ into a sparse matrix using one-hot encoding, and append time, $t$, as a stand-alone variable to create the feature vector. 
$\pi^{pred}$ is the target variable and, using this model, a query's estimated arrival at $v_d$ is:
\begin{equation}
    \textstyle
    \label{eq:expected-journey-time}
    \xi_{sdt} = a_s + |\phi_{sdt}| + \pi^{pred}_{sdt}
\end{equation}

\para{Minimizing Redundant Computations}
Once a path has been assigned, only paths that have edge-time locations in common with the assigned path (i.e., where ELM values change) will experience changes in arrival time. 
Hence, CS-MAT does not re-compute all paths in $\mathcal{Q}'$. It is sufficient to only re-process paths that intersect with the previously assigned path.

\para{Rolling Query Batches}
CS-MAT can also be performed by using rolling batches within $\mathcal{Q}$. 
A pre-specified time parameter, $y$, can control how `far' into $\mathcal{Q}$ to search for candidate queries (measured from the departure time of the current query).
This restricts the size of $\mathcal{Q}'$, which may be useful in low CPU environments. 

\subsubsection{Design Details}
We now proceed to outline CS-MAT, as described in Algorithm \ref{alg:CS-MAT}.

\para{Batch Forming}
After initializations (Lines 1-2), CS-MAT selects the next chronological query in $\mathcal{Q}$ from which to define $\mathcal{Q}_{batch}$. 

The algorithm forms $\mathcal{Q}_{batch}$ from the queries entering the system in a time window, controlled by $y$.
The unprocessed queries in $\mathcal{Q}_{batch}$ are sorted by their pre-computed free-flow arrival time (Line 5) and we then iterate through $\mathcal{Q}_{batch}$ (from Line 7).

\para{Simple Case}
The goal at each iteration is to find the unprocessed query in $\mathcal{Q}_{batch}$ that returns the earliest arrival time, given the expected future load. 
In the simplest case, the first query (Line 15) is tested to see if any edges are in a congested regime (Lines 11-14). 
If no edges are, we know this query must have the earliest arrival time in $\mathcal{Q}_{batch}$ (as $\mathcal{Q}_{batch}$ is already sorted by free-flow arrival time) and so we assign its path, update the ELM, remove the query from $\mathcal{Q}_{batch}$, and move to the next query (Lines 15-18).

\para{Restricting the Search Space}
If the first query has any congested edges, we need to define $\mathcal{Q}'$, which is the set of candidate queries from which to find the earliest arrival time.
There are two sub-cases (controlled by \textit{reCheck}).
In sub-case 1 (Lines 20-22), we encounter a `base' query for the first time and need to find all possible candidate queries.
To determine $\mathcal{Q}'$, the expected arrival time is calculated, $\xi_{sdt}$ of the current $q_{sdt}$.
To populate $\mathcal{Q}'$, all queries in $\mathcal{Q}_{batch}$ with an earlier free-flow arrival time than the expected arrival time are selected.
In sub-case 2 (Lines 23-24), $\mathcal{Q}'$ is already known (e.g., the same base query is being re-processed) and the queries therein already calculated. 
We therefore only want to identify those queries in $\mathcal{Q}'$ that need to be recalculated.
The calculated paths in $\mathcal{Q}'$ are recorded in a path matrix $P(e,\tau)$, where path ID is recorded with its $(e,\tau)$ tuple. 
Next, $\mathcal{Q}'$ is updated by redefining it as the set of all the queries whose ID intersects with previously assigned path ID.  
That is, all $q' \in \mathcal{Q}'$ for which at least one edge in $p'$ intersects, both spatially and temporally, with the previously assigned path.

\para{Determining Temporal Load-Aware Shortest Paths}
Once $\mathcal{Q}'$ has been defined, the temporal load-aware paths for each $q'$ is determined using TLAA* and added to $P(e,\tau)$ (Lines 27-28). 
Each path can be determined in parallel without any loss of information. 
The returned paths are compiled and the path with the earliest arrival time is assigned (Lines 29-31).  
Finally, we update $\mathcal{P}$ with this best path, update $\mathcal{L}(\mathcal{E},\tau)$, and remove the corresponding query from both $\mathcal{Q}$ and $\mathcal{Q}_{batch}$ (Lines 32-35).

\begin{algorithm}[t]
\small
\algrenewcommand\algorithmicindent{0.75em}%
	\caption{Collective Search For Minimal Arrival Time (CS-MAT)}
	\label{alg:CS-MAT}
	\begin{algorithmic}[1]
	\Require{$\mathcal{G}(\mathcal{V},\mathcal{E})$, $\mathcal{Q}$,$y$}
	\Ensure{$\mathcal{P}$}

    \State Initialize $\mathcal{L}(\mathcal{E},\tau)$
    \State $\mathcal{P}=\emptyset$
    
    \While{$\mathcal{Q} \neq \emptyset$}
    
        \State Define $\mathcal{Q}_{batch}$ \Comment{Determined using $y$}
        \State Sort $\mathcal{Q}$ by free-flow arrival time
        \State $reCheck \gets False$
        
        \While{$\mathcal{Q}_{batch} \neq \emptyset$}
        
            \If{$reCheck = False$}
                \State $q_{sdt} \gets$ Next query in $\mathcal{Q}_{batch}$
            \EndIf

            \State $congPath \gets False$
            \ForAll{$e_{i,j} \in \phi_{sdt}$} 
    	        \If{$l(e,\tau) \geq F_{ij}$}
    	            \State $congPath \gets True$
    	            \Break
    	        \EndIf
            \EndFor
            
            \If{$congPath = False$} \Comment{Free-flow best path available}
                \State $\mathcal{P}  \gets \mathcal{P} \cup \phi_{sdt}$
                \State Update $\mathcal{L}(\mathcal{E},\tau)$
                \State $\mathcal{Q}_{batch} \gets \mathcal{Q}_{batch} \setminus q_{sdt}$
                
            \Else \Comment{Best path congested}
                \If{$reCheck = False$}
                    \State Initialize $P(e,\tau)$ \Comment{Path-Edge Matrix}
                    \State Define $\mathcal{Q}'$ \Comment{(Sub-case 1)}
                \Else 
                    \State Define $\mathcal{Q}'$ \Comment{(Sub-case 2)}
                \EndIf
            
                \State $p^* \gets \emptyset;\quad q^* \gets \emptyset;\quad a_{sdt}^* \gets \infty$
                \ForAll{$q' \in \mathcal{Q}'$}
                    \State $p' \gets$ \textsc{TLAA*}$(\mathcal{G}(\mathcal{V},\mathcal{E}), \mathcal{L}(\mathcal{E},\tau), q')$
                    \State Add $p'$ to $P(e,\tau)$
                    \State Calculate $a'_{sdt}$
                    \If{$a'_{sdt} < a_{sdt}^*$}
                        \State $p^* \gets p';\quad q^* \gets q';\quad a_{sdt}^* \gets a'_{sdt}$
                    \EndIf
                \EndFor
                
                \State $\mathcal{P} \gets \mathcal{P} \cup p^*$
                \State Update $\mathcal{L}(\mathcal{E},\tau)$
                \State $\mathcal{Q} \gets \mathcal{Q} \setminus q'$
                \State $\mathcal{Q}_{batch} \gets \mathcal{Q}_{batch} \setminus q'$
                \State Update $reCheck$
            \EndIf
        \EndWhile
    \EndWhile
    \State \textbf{return} $\mathcal{P}$
	\end{algorithmic}
\end{algorithm}

\subsubsection{Scalability}
\label{sss:cs-mat-efficiency}
CS-MAT enables query processing to be embarrassingly parallel. 
When the shortest path from one query in $\mathcal{Q}'$ is determined, it does not affect the processing of another. 
We can achieve this while still being collective because the shortest path calculations within a single iteration are only dependent on the ELM, rather than one another. 
Once a path has been assigned its impact on expected system load, and on other queries, in the candidate set $\mathcal{Q}'$ is accounted for in subsequent iterations. 
After batching, each iteration has two operations. 
The `outer operation' selects the candidate query set, prepares it for parallelization, and then recompiles the shortest path results to identify which query in $\mathcal{Q}'$ to assign.
The `inner operation' calculates the shortest paths and can use any shortest path algorithm (e.g., TLAA*).
The complexity is upper-bounded by $\mathcal{O}(Q (V\log V + ET))$ due to the maximum number of outer iterations (in practice far fewer than $Q$). 
The inner cost is the complexity of TLAA*.
We observe that the runtime of the outer operation is typically in the order of milliseconds, and always less than half a second in our experimental setting. 
Hence, when fully parallelized, the actual algorithm runtime is not much more than that of a stand-alone shortest path calculation.

\section{Experiments}
\label{s:experiments}
This section outlines our experimental setting, the datasets, performance metrics, and baselines.
The experiments use representative query workloads as the basis for travel demand, which mirrors common practice in real-life traffic management in which OD matrices are used as the basis for demand information.

\subsection{Modeling Road Networks}
We integrate a number of concepts that are prevalent in the traffic management community, including traffic flow regimes \cite{wang2009speed,may1990traffic}, safe headway \cite{highwaycode_headway}, and transition penalties \cite{brackstone1999car}. 
The base headway, $\eta$, is the number of seconds required between vehicles for them to safely traverse an edge.
A transition penalty, $\psi$, is incurred when moving from one road to another, and this captures elements of travel such as waiting at traffic lights or slowing to turn a corner.
The free-flow capacity, $F_{ij}$, is the maximum number of vehicles that can safely traverse an edge at the speed limit in a given time interval. 
We model this as the number of vehicles that can exist on the edge while maintaining safe headway distance, plus the number of vehicles that can safely join the edge in the given time interval $\tau$. 
Formally, it is given as: 
\begin{equation}
    \textstyle
    \label{eq:max-free-flow}
    F_{ij} = \frac{\delta_{ij}}{z_{ij} \times \eta} + \frac{I}{\eta + \psi}
\end{equation}

\para{Assumptions}
To establish a generalizable basis for collective routing, we assume all queries result in a journey, all users begin their journeys as soon as their path is given to them, and all users follow the assigned path. FIFO compliance means there is no overtaking in the network.
In determining $F_{ij}$, we assume that all vehicles are of equal size.
This assumption can be relaxed to account for other vehicle types (e.g., trucks, buses) by following best practice methods from traffic modeling literature.

\subsection{Data}
\label{ss:real-world}
We use real taxi trip data from two cities: Porto, Portugal \cite{Porto2015} and New York, USA \cite{NYC2020}. 
The cities have contrasting topologies: New York has an orderly grid-based road network, whereas Porto has a much less orderly network.

We extract 3km $\times$ 3km regions from OpenStreetMap (OSM) using \texttt{osmnx} \cite{Boeing2017}. 
These regions are centered on busy sites that show temporal variation: Penn Station for New York, and S\~{a}o Bento Station for Porto. 
Selecting these regions enable us to cover most of Porto's city center, and a significant proportion of Manhattan in New York.
We apply geographical and temporal filters: we only select trips that start and end within these regions, and that start in the range 7-11am on a weekday (i.e., to simulate the rush-hour). 
We assume all journeys take place on the same day. 
We take edge attributes from OSM (and make inferences where data is unavailable). 

As Porto is a smaller city than New York, we use smaller query sets for Porto (2k, 5k, 10k, and 25k), and larger query sets in New York (10k, 25k, 50k, and 100k) to represent their real-world congestion levels. For example, millions of journeys occur daily across New York, so 100k trips in a 9km\textsuperscript{2} region across a four-hour window can be broadly representative of very high congestion.

\subsection{Evaluation Measures}
We evaluate the methods using the following measures.
Average journey time (AJT), in minutes, is defined as:
\begin{equation}
    \textstyle
    AJT =  \frac{1}{P}\sum_{p \in \mathcal{P}} |p_{sdt}|
\end{equation}
Free-flow capacity utilization (FFCU) assesses what proportion of the free-flow capacity in the network has been utilized.  To define FFCU, we first define $\sigma(e_{ij}, \tau)$, which represents the proportion of free-flow capacity that is utilized along an individual edge:
\begin{equation}
    \textstyle
    \label{eq:sigma}
    \sigma(e_{ij}, \tau) = 
        \begin{cases}
            \frac{l(e_{ij}, \tau)}{F_{ij}}  & \text{if } l(e_{ij}, \tau) \leq F_{ij}\\
            1 & \text{if } l(e_{ij}, \tau) > F_{ij} \\
        \end{cases}
\end{equation}
FFCU is then defined as:
\begin{equation}
    \textstyle
    FFCU = \frac{1}{ET} \sum_{\tau \in \mathcal{T}}\sum_{e_{ij} \in \mathcal{E}} \sigma(e_{ij}, \tau)
\end{equation}
Load distribution (LD) returns the proportion of edges that are being utilized (i.e., have a non-zero load), defined as:
\begin{equation}
    \textstyle
    LD = \frac{1}{ET} \sum_{\tau \in \mathcal{T}} \sum_{e_{ij} \in \mathcal{E}} \rho(e_{ij}, \tau)
\end{equation}
where $\rho(e_{ij}, \tau) = 1$ if $l(e_{ij}, \tau)$ is non-zero, and zero otherwise.

We evaluate the fairness of a given solution by assessing the distribution of each user's congestion penalty. 

\subsection{Experimental Set Up}
\label{ss:expt-set-up}
In our experiments, we evaluate TLAT$k$, TLAA*, and CS-MAT. To process queries, we sort $\mathcal{Q}$ in time-ascending order (i.e., those queries that start earliest are processed first). 
CS-MAT does not require the query set to be (time-)sorted as it is a collective solution.

The congestion alleviation methods used by Google Maps, TomTom, etc. are proprietary. 
However, we model what we understand from the traditional practice which is to route vehicles based on the currently known load distribution when the query is asked \cite{jeong2015saint}, but without explicit consideration for the future impact of that routing.
We call this method Static Load-Aware Dijkstra (SLAD) and it forms one of our baselines. 
As TLAA* calculates transitions through the network using a load-aware arrival time function, so does SLAD except the underlying ELM reflects only the static load distribution at the timestep in which the query is initiated. 
It can therefore only adapt to congestion known at the point at which the query is asked.
We also implement Free-Flow Na\"{i}ve Dijkstra (FFND) as another baseline, which assigns each path its free-flow best path given by the basic version of Dijkstra's shortest path algorithm \cite{dijkstra1959note}. 
In assessing both of these baselines, we calculate the actual path cost using the actual load within the system and the arrival time function.
We do not compare the methods presented in \cite{motallebi2020streaming} as these depend on historical road network data, which is unavailable to us, or \cite{2020-sub-28} as their optimization approach is fundamentally different.

Finally we set:
$k = 5$, $I$ = 360s, $\eta$ = 3s \cite{fairclough1997effect}, and $\psi = 0.5\Upsilon_{ij}$ seconds \cite{levinson_2018}. 
Within CS-MAT, $y$ is set to 14,400s (=4 hours), which means all queries are ingested as a single batch.
The algorithms are implemented in Python on a Linux machine with 64GB RAM and six CPU cores (all of which are used in our experiments).

\section{Results}
\label{s:results}

We first analyze the performance of our temporal load-aware shortest path algorithms, then we evaluate the effect of computing shortest paths collectively using CS-MAT.
We also investigate scenarios in which only a portion of the vehicles are routed collectively. 
Our analysis initially focuses on average journey times, and we then consider other aspects, such as network utilization and fairness. 

\subsection{Temporal Load-Aware Algorithms}

Temporal load-aware adaptations of popular algorithms are the `engine' of any collective query processing, including our solution -- CS-MAT. 
They can also be used as stand-alone solutions within existing navigational systems by modeling environments as TLANs.

\begin{figure}[t]
    \centering
    \includegraphics[width=0.98\columnwidth]{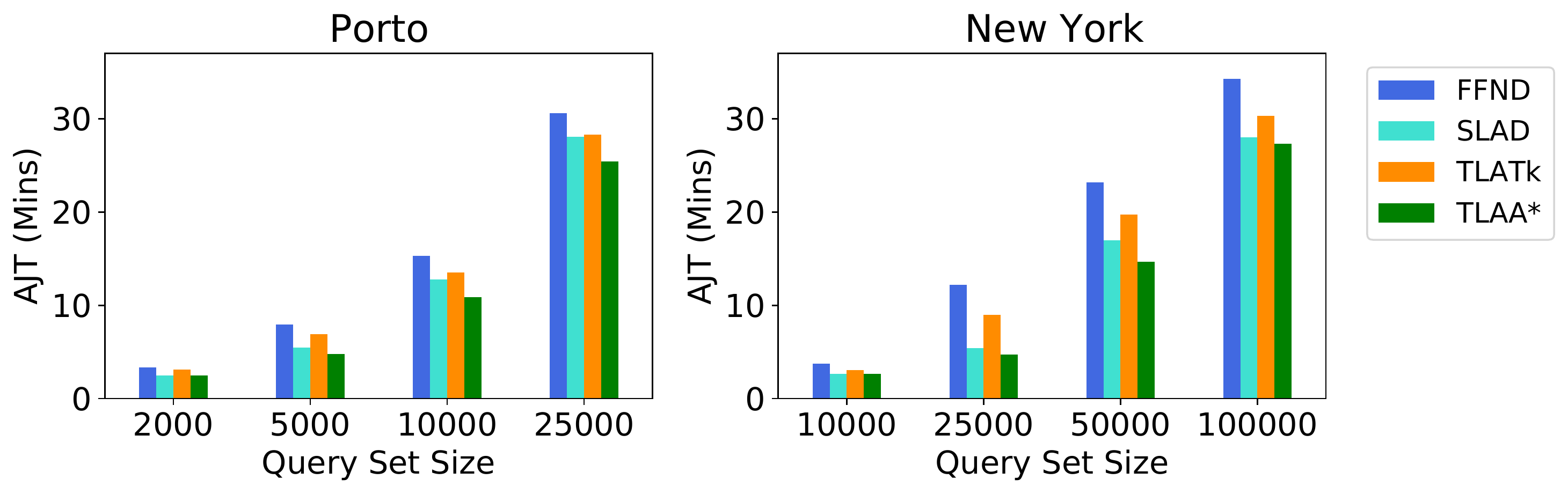}
    \caption{AJT of temporal load-aware algorithms}
    \label{fig:algs-ajt}
\end{figure}

Figure \ref{fig:algs-ajt} shows that both TLAT$k$ and TLAA* report a lower AJT than FFND, and that TLAA* outperforms SLAD. 
Overall, TLAA* is consistently the best performing algorithm for computing the shortest paths in TLANs. 
We can evaluate the impact of operating in a TLAN by comparing SLAD and TLAA*. 
It is consistently beneficial to consider expected future traffic load, and the benefit tends to increase in more congested systems.
However, under very high congestion, the difference in performance starts to plateau (e.g., in New York, there is only a 2.5\% improvement under very high congestion, as opposed to 13.6\% in high congestion regimes).
This is expected as it is naturally harder to find free-flow capacity, or even viable alternative routes, when a significant number of edges are already heavily congested. 
Actual time savings are also strong: in New York, TLAA* saves 8.5 and 6.9 minutes (37\% and 20\%) compared to FFND when congestion is high and very high respectively, and 2.3 and 0.7 minutes compared to SLAD (15.7\% and 2.5\%).

\subsection{Collective Query Processing}
\label{ss:query-processing}
\begin{figure}[t]
    \centering
    \includegraphics[width=0.98\columnwidth]{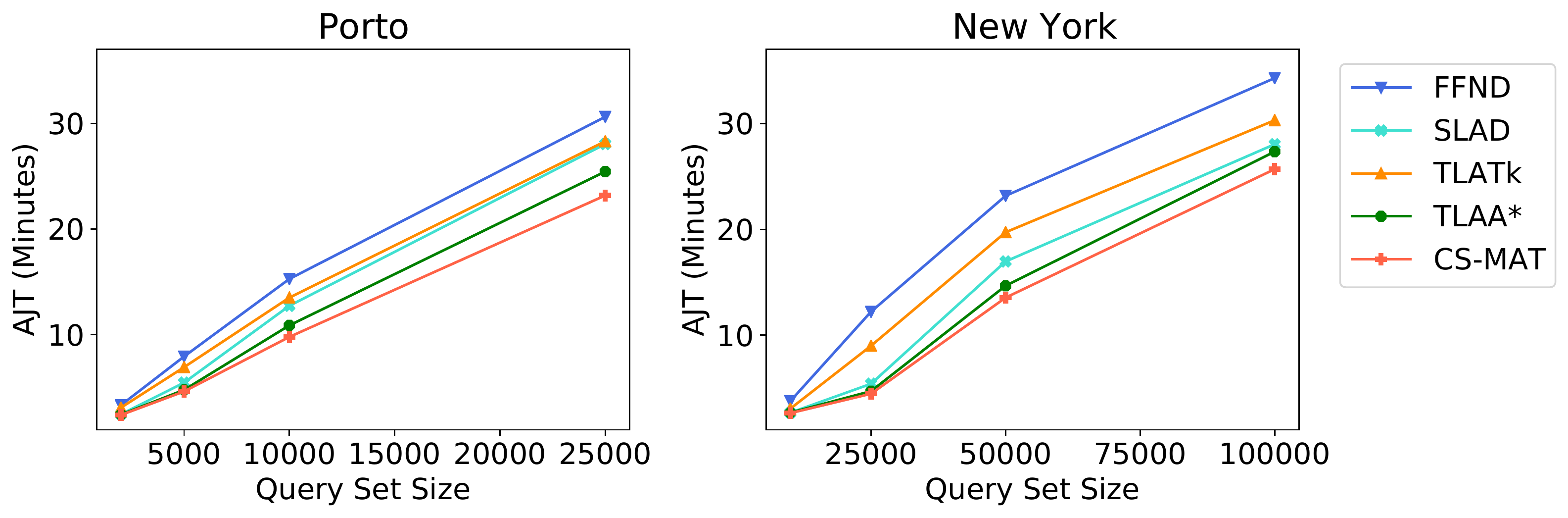}
    \caption{AJT across varying query set sizes}
    \label{fig:heuristic-results}
\end{figure}

Comparing CS-MAT with other solutions indicates the impacts of collectively routing.
Figure \ref{fig:heuristic-results} shows that CS-MAT is the most effective algorithm at reducing AJTs in all scenarios. 
In Porto, CS-MAT improves upon TLAA* by 11.2\% and 9.8\% in high and very high congested systems, respectively.
In New York, CS-MAT is 8.3\% and 6.5\% better than TLAA* in the same congestion regimes.
As congestion increases, the benefit of collective processing becomes clearer, which is to be expected as more edges are likely to be over-capacitated and more re-routing is necessary.

We also compare the performance of CS-MAT with the baselines. 
Compared to FFND, CS-MAT significantly improves AJT; in Porto, CS-MAT creates AJT savings of up to 7.4 minutes in systems with high congestion, and we also see significant savings in medium congestion (42\%, giving a saving of 5.5 minutes). 
In New York, CS-MAT makes bigger savings compared to FFND, where AJT is reduced by up to 63.5\% (medium congestion).
Against SLAD, CS-MAT is saving 2.9 and 4.9 minutes in high and very high congestion systems in Porto (23.1\% and 17.3\%), and 3.4 and 2.4 minutes in New York (20.2\% and 8.4\%). 
These time savings aggregate significantly at the system level (e.g., in New York, an AJT saving of three minutes for each of the 50k users gives a system-level saving of 2,500 hours).

In less-congested scenarios, the benefit of using CS-MAT over TLAA* is less pronounced. 
In a low congestion system, any algorithm is likely to return the free-flow best path for a query, as such ordering of queries matters less as the distribution of load in the system is less likely to adversely impact travel times for other routes. 
Where CPU capacity is limited, TLAT$k$ may be considered as it performs well against ND as it is still able to make some dynamic decisions even if these are limited to the $k$ pre-computed paths.
\begin{figure*}[t]
    \includegraphics[width=\textwidth]{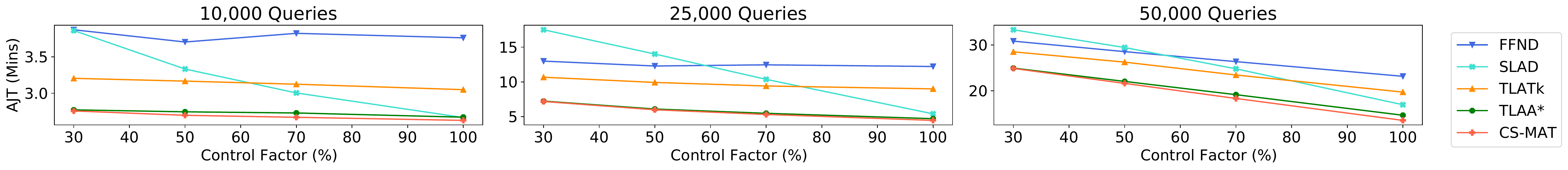}
    \caption{Variation in AJT in New York as the control factor varies; same legend in all sub-figures}%
    \label{fig:uncontrolled-load-nyc}%
\end{figure*}

\subsection{Uncontrolled Load}
\label{ss:uncontrolled-load}
While modeling and managing travel demand are attracting attention in modern traffic management, there are still limitations today in terms of monitoring the traffic loads.
As autonomous/assisted driving is not yet widespread, we cannot expect every user to have their route assigned according to the same collective goal. 
Hence, we evaluate our algorithms under lower levels of load control. 
We generate a set of representative queries and process these using TLAA*. 
Where we reduce our level of control, we calculate a `base' load in the system based off of these representative ELMs. 
The loads are reduced by $\gamma$\%, where $\gamma$ is a control factor. 
For example, if an edge has a predicted load of 10, the uncontrolled load would be 7 for 30\% control, 5 for 50\% control, etc.

Figure \ref{fig:uncontrolled-load-nyc} shows that CS-MAT remains the best performing algorithm in terms of AJT. 
While the performance difference between CS-MAT and TLAA*/FFND tends to reduce at lower levels of control, the difference compared to SLAD actually increases with reduced levels of control which is an important finding. 
Not only is collective and dynamic routing effective even at lower levels of traffic control, but by not doing so when other operators are, existing approaches can actually return even worse results at lower levels of control. 
This is because the traffic load experienced by other operators is no longer as predictable as it once was, due to the TLA systems that dynamically respond to and re-route traffic.

\subsection{Network Utilization}
\label{ss:utilization}
Recent studies \cite{2020-sub-24} indicate that utilizing spare capacity can be an inexpensive way of reducing congestion and positively impacting quality of life and air pollution.  
Accordingly, we consider how collective routing affects network utilization and load distribution.

\begin{figure}[t]
    \centering
    \begin{subfigure}[b]{\columnwidth}
        \includegraphics[width=\textwidth]{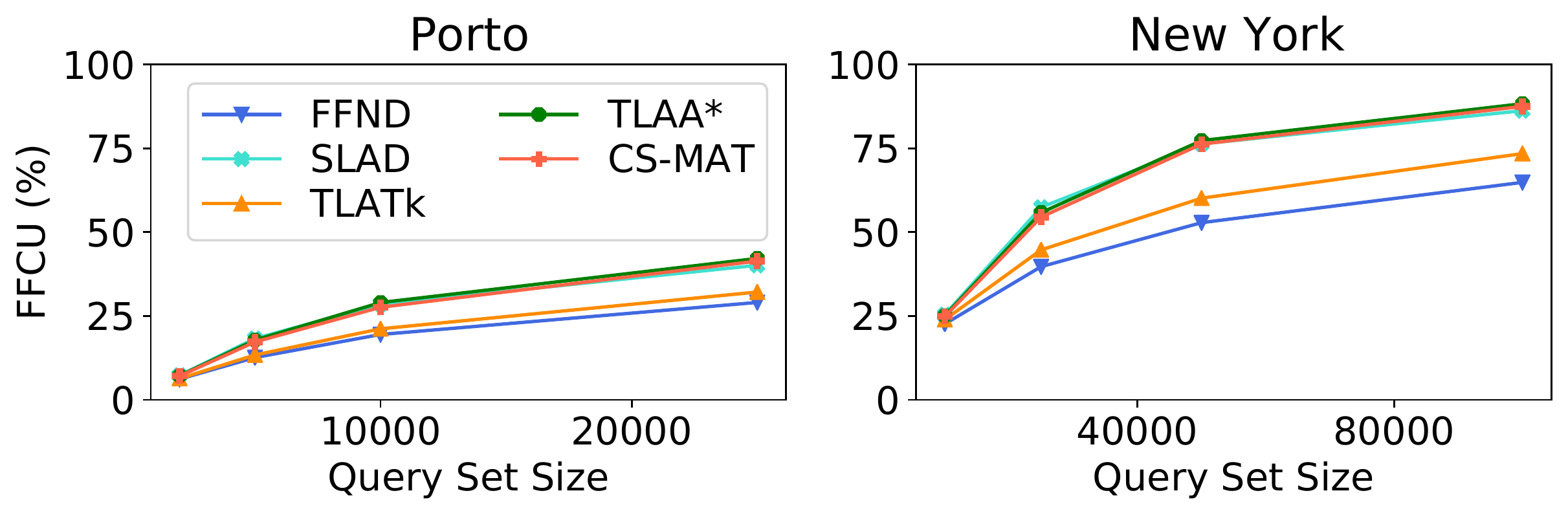}
        \caption{Free Flow Capacity Utilization}
        \label{fig:util-ffcu}
    \end{subfigure}
    \\
    \begin{subfigure}[b]{\columnwidth}
        \includegraphics[width=\textwidth]{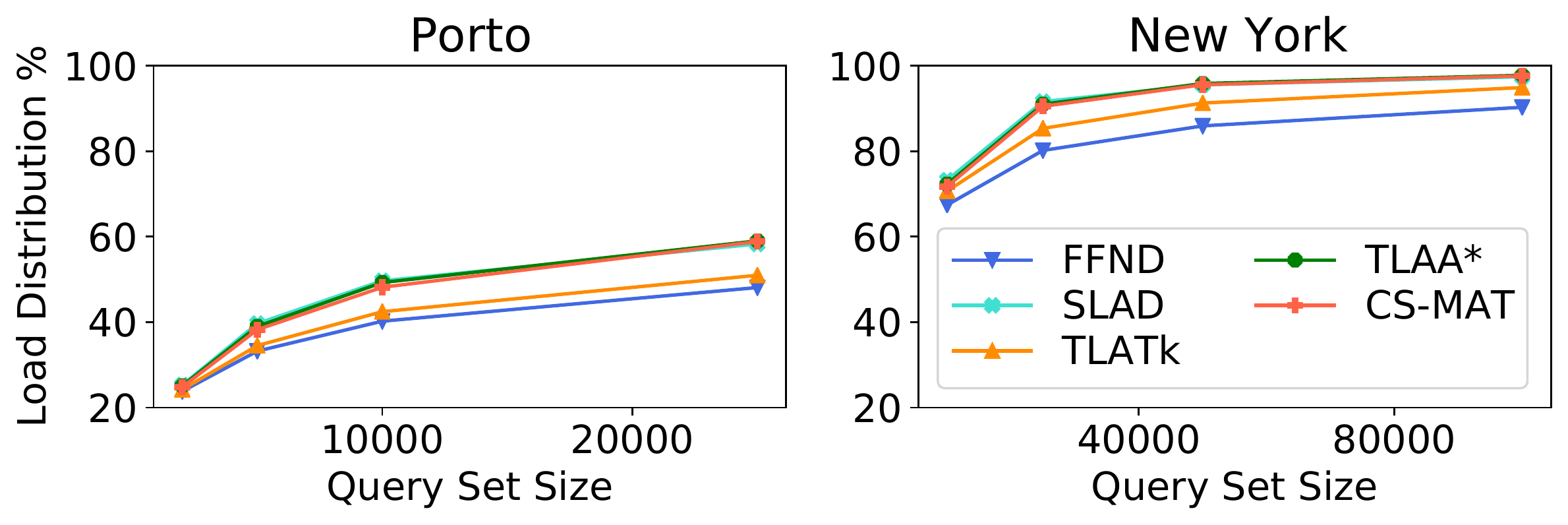}
        \caption{Load Distribution}
        \label{fig:util-ld}
    \end{subfigure}
    \caption{Network utilization for Porto and New York; same legend in all sub-figures}%
    \label{fig:algo-results-ny-porto-congestion}%
\end{figure}

Figure \ref{fig:util-ffcu} shows that CS-MAT and TLAA* are able to find uncongested edges better than the baselines. 
We expect this as these algorithms are making dynamic edge-level decisions driven by the arrival time function, which penalizes congested edges. 
The difference in performance compared to FFND and TLAT$k$ increases with mid-to-high levels of congestion, before plateauing when congestion intensifies. 
FFCU improves by 42\% and 30\% compared to FFND and TLAT$k$ in Porto in a highly congestion system (44\% and 27\% for New York). 
There is little difference between CS-MAT and TLAA*; TLAA* is slightly better at finding capacity, despite generally reporting slower journey times. 
This indicates there exists a trade-off between finding capacity and reducing AJT.  
That is, at some point, it is no longer beneficial to find free-flow capacity for its own sake.

Figure \ref{fig:util-ld} shows that TLAA*, CS-MAT, and SLAD distribute load across a more diverse set of edges than TLAT$k$ and FFND, and the difference in performance increases in mid-to-high congestion level systems, but plateaus somewhat with very high congestion. 
In New York, although query set sizes are high, it is notable that CS-MAT is able to reach up to 97\% load distribution, which is far higher than in Porto.
The performance improvements achieved by our algorithms in New York compared to Porto emphasizes the importance of topology. 
While this may highlight the benefits of planned cities with organized, uniform road networks, it also indicates further research potential for algorithms that are specifically designed to find capacity within less-ordered networks.

\subsection{Fairness}
\label{ss:fairness}
Inherent to any collective solution is that some individuals will be assigned paths that are not their theoretical free-flow shortest path. 
While we attempt to optimize journey times at the system level, it is important to understand how fair each system is for individuals.

\begin{figure}[t]
    \centering
    \includegraphics[width=\columnwidth]{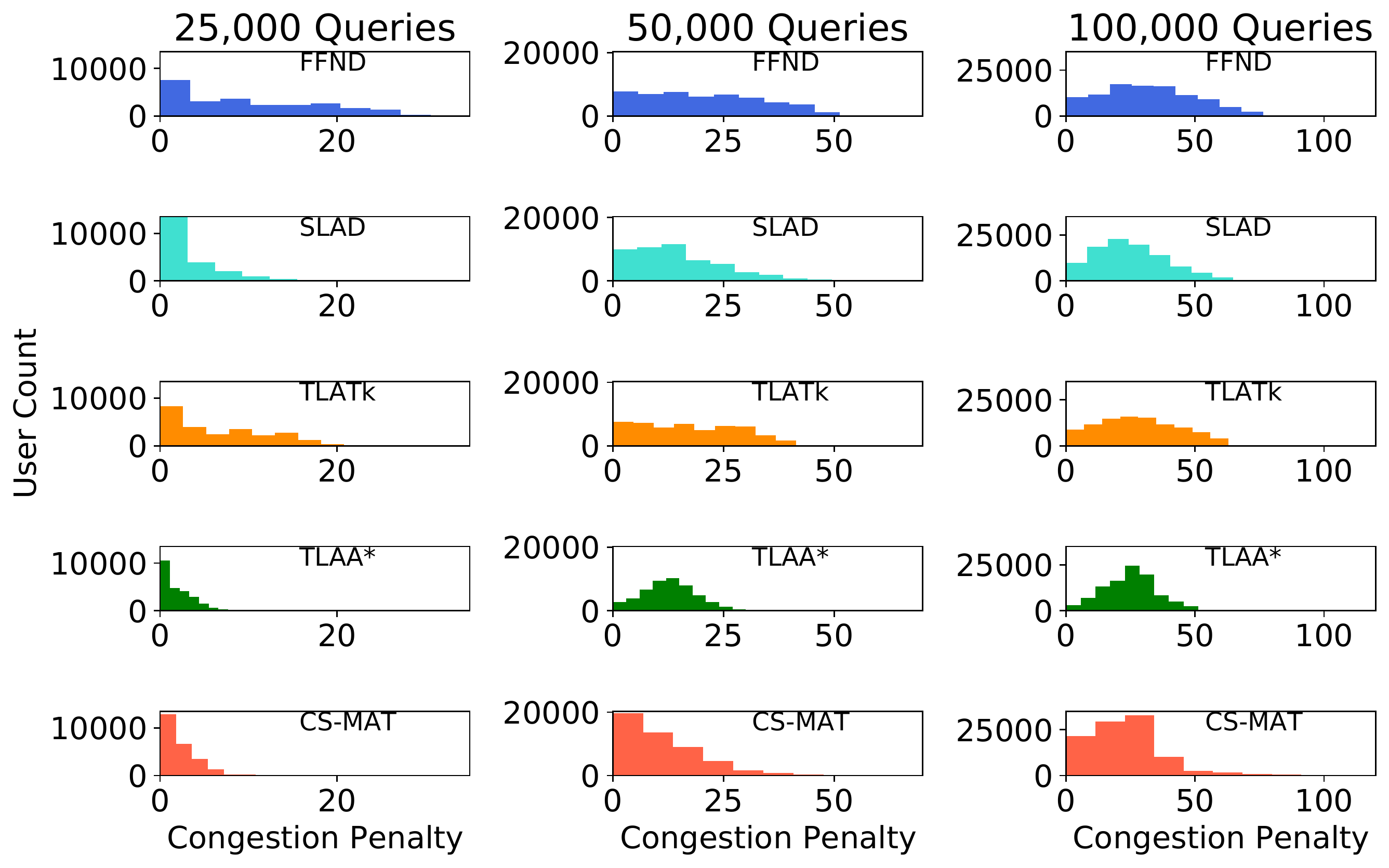}
    \caption{Distribution of congestion penalties (New York)}
    \label{fig:fairness-hists}
\end{figure}

Figure \ref{fig:fairness-hists} shows the distribution of the congestion penalties (in minutes). 
Due to space limitations, we only show results for New York; Porto exhibits similar patterns. 
When congestion is high, CS-MAT has a wider spread of penalties while showing a positive skew, whereas TLAA* has a smaller spread and we observe a tendency towards a Gaussian distribution, particularly in higher congestion systems.
Comparing TLAA* to CS-MAT under high congestion, the mean penalty for CS-MAT is lower (TLAA*: 12.2, CS-MAT: 11.1), but TLAA* has lower standard deviation (TLAA*: 5.9, CS-MAT: 9.6). 
TLAA* tends to offer a fairer distribution as fewer individuals receive `harsh' penalties and, with a lower standard deviation, most users can expect similar penalties to others in the system.
This indicates a trade-off may exist between reducing AJT and ensuring fairness in the system; this warrants further study.

\subsection{Runtime Analysis}
We also analyze algorithm runtime within the context of AJT improvements. 
For this, we measure the `end-to-end' time, which is the time taken for the algorithm to return a route, and for the journey to be completed. 
We parallelized the execution across six nodes, which is typically fewer than would be available in a real-world deployment of any solution. 

Formally, the end-to-end time is the sum of the `per query' algorithm runtime (e.g., processing time) and the AJT.  
We take the average over the query set, and present these values in Table \ref{tab:nyE2Etime}.  
CS-MAT is the best-performing algorithm for all levels of congestion, although TLAA* performs well. 
These results demonstrate the potential benefit of collective processing in providing socially conscious routing.
With sufficient computing power and optimized implementations of the underlying shortest path algorithm, the collective approach can be more effective and significant additional time savings are feasible.
\section{Conclusion}
\label{s:conclusion}

\begin{table}[t]
\caption{End-to-end time, in minutes (New York)}
\label{tab:nyE2Etime}
\begin{tabular}{cccccc} 
\toprule
\textbf{No. Queries} & {\textbf{ND}} & {\textbf{SLAD}}&{\textbf{TLAT\textit{k}}} & {\textbf{TLAA*}} & {\textbf{CS-MAT}} \\
\midrule
10,000  & 3.76 & 2.67 & 3.05 & 2.67 & \textbf{2.63} \\
25,000  & 12.21 & 5.40 & 8.99 & 4.70 & \textbf{4.50} \\ 
50,000  & 23.16 & 16.96 & 19.72 & 14.66 & \textbf{13.95} \\ 
100,000 & 34.28 & 28.04 & 30.30 & 27.34 & \textbf{26.12} \\
\bottomrule
\end{tabular}
\end{table}

In this paper, we addressed the problem of collectively processing shortest path queries with the aim of minimizing system-wide congestion. 
To solve the CP-SPQ problem, we formalized the temporal load-aware setting, in which the expected future load in the network is tracked, and an edge-level arrival time function penalizes congested edges. 
When queries are answered, and paths assigned, the network's tracking matrices are updated and future queries can take account of this expected load within the system. 

We presented adaptations of classic algorithms to operate in TLANs.
TLAA* adapts A*, enabling it to utilize future traffic states in order to dynamically and proactively avoid congestion thus finding alternative routes. 
TLAT$k$ pre-computes the $k$ best free-flow paths, dynamically calculates the cost of these paths given current load in the system, and selects the best path for the query. 
Our solutions are more effective (i.e., they return a lower AJT) and can be as efficient as the baselines if fully parallelized.
Our work is not just for the future; it would be beneficial for any navigational service to model their environment in a temporal load-aware way, and adapt their existing algorithms to be temporal- and load-aware. 
Experiments show that this can be effective even when operators have access to only a proportion of the traffic flow.

Finally, we introduced CS-MAT, which considers queries in a collective manner, by iteratively searching for the lowest arrival time from a query batch. 
We used machine learning and designed data types to track processing results between iterations 
in order to minimize redundant calculations, and discussed how CS-MAT is embarrassingly parallel, which indicates its potential for large-scale deployment.
Not only does CS-MAT reduce AJTs for individuals by up to 20\% in New York and 23\% in Porto compared to non-collective solutions, it also better utilize the available capacity in the network finding up to 44\% more free-flow capacity.
These results emphasize the potential benefit that collectively processing shortest path queries on road networks can have on urban life.

\begin{acks}
This work is supported in part by the \grantsponsor{epsrc}{UK Engineering and Physical Sciences Research Council} u under Grant No. \grantnum{epsrc}{EP/L016400/1}.
We also thank Elif Eser for her work in the early stages of the project.
\end{acks}


\bibliographystyle{ACM-Reference-Format}
\bibliography{99-bib.bib}


\begin{thebibliography}{34}


\ifx \showCODEN    \undefined \def \showCODEN     #1{\unskip}     \fi
\ifx \showDOI      \undefined \def \showDOI       #1{#1}\fi
\ifx \showISBNx    \undefined \def \showISBNx     #1{\unskip}     \fi
\ifx \showISBNxiii \undefined \def \showISBNxiii  #1{\unskip}     \fi
\ifx \showISSN     \undefined \def \showISSN      #1{\unskip}     \fi
\ifx \showLCCN     \undefined \def \showLCCN      #1{\unskip}     \fi
\ifx \shownote     \undefined \def \shownote      #1{#1}          \fi
\ifx \showarticletitle \undefined \def \showarticletitle #1{#1}   \fi
\ifx \showURL      \undefined \def \showURL       {\relax}        \fi
\providecommand\bibfield[2]{#2}
\providecommand\bibinfo[2]{#2}
\providecommand\natexlab[1]{#1}
\providecommand\showeprint[2][]{arXiv:#2}

\bibitem[\protect\citeauthoryear{Angelelli, Morandi, Savelsbergh, and
  Speranza}{Angelelli et~al\mbox{.}}{2021}]%
        {angelelli2016system}
\bibfield{author}{\bibinfo{person}{Enrico Angelelli},
  \bibinfo{person}{Valentina Morandi}, \bibinfo{person}{Martin Savelsbergh},
  {and} \bibinfo{person}{Grazia Speranza}.} \bibinfo{year}{2021}\natexlab{}.
\newblock \showarticletitle{System optimal routing of traffic flows with user
  constraints using linear programming}.
\newblock \bibinfo{journal}{\emph{European Journal of Operational Research}}
  \bibinfo{volume}{293}, \bibinfo{number}{3} (\bibinfo{year}{2021}),
  \bibinfo{pages}{863--879}.
\newblock


\bibitem[\protect\citeauthoryear{Barth and Funke}{Barth and Funke}{2019}]%
        {2020-sub-33}
\bibfield{author}{\bibinfo{person}{Florian Barth} {and} \bibinfo{person}{Stefan
  Funke}.} \bibinfo{year}{2019}\natexlab{}.
\newblock \showarticletitle{Alternative Routes for Next Generation Traffic
  Shaping}. In \bibinfo{booktitle}{\emph{ACM SIGSPATIAL}}.
  \bibinfo{pages}{1--8}.
\newblock


\bibitem[\protect\citeauthoryear{Boeing}{Boeing}{2017}]%
        {Boeing2017}
\bibfield{author}{\bibinfo{person}{Geoff Boeing}.}
  \bibinfo{year}{2017}\natexlab{}.
\newblock \showarticletitle{OSMnx: New methods for acquiring, constructing,
  analyzing, and visualizing complex street networks}.
\newblock \bibinfo{journal}{\emph{Computers, Environment and Urban Systems}}
  \bibinfo{volume}{65} (\bibinfo{year}{2017}), \bibinfo{pages}{126--139}.
\newblock


\bibitem[\protect\citeauthoryear{Brackstone and McDonald}{Brackstone and
  McDonald}{1999}]%
        {brackstone1999car}
\bibfield{author}{\bibinfo{person}{Mark Brackstone} {and} \bibinfo{person}{Mike
  McDonald}.} \bibinfo{year}{1999}\natexlab{}.
\newblock \showarticletitle{Car-following: a historical review}.
\newblock \bibinfo{journal}{\emph{Transportation Research Part F: Traffic
  Psychology and Behaviour}} \bibinfo{volume}{2}, \bibinfo{number}{4}
  (\bibinfo{year}{1999}), \bibinfo{pages}{181--196}.
\newblock


\bibitem[\protect\citeauthoryear{Buzna and Carvalho}{Buzna and
  Carvalho}{2017}]%
        {2020-sub-24}
\bibfield{author}{\bibinfo{person}{Lubo{\v{s}} Buzna} {and}
  \bibinfo{person}{Rui Carvalho}.} \bibinfo{year}{2017}\natexlab{}.
\newblock \showarticletitle{Controlling congestion on complex networks:
  fairness, efficiency and network structure}.
\newblock \bibinfo{journal}{\emph{Scientific reports}} \bibinfo{volume}{7},
  \bibinfo{number}{1} (\bibinfo{year}{2017}), \bibinfo{pages}{1--15}.
\newblock


\bibitem[\protect\citeauthoryear{Cheng, Gkountouna, Z{\"u}fle, Pfoser, and
  Wenk}{Cheng et~al\mbox{.}}{2019}]%
        {cheng2019shortest}
\bibfield{author}{\bibinfo{person}{Dan Cheng}, \bibinfo{person}{Olga
  Gkountouna}, \bibinfo{person}{Andreas Z{\"u}fle}, \bibinfo{person}{Dieter
  Pfoser}, {and} \bibinfo{person}{Carola Wenk}.}
  \bibinfo{year}{2019}\natexlab{}.
\newblock \showarticletitle{Shortest-Path Diversification through Network
  Penalization: A Washington DC Area Case Study}. In
  \bibinfo{booktitle}{\emph{ACM SIGSPATIAL IWCTS}}. \bibinfo{pages}{1--10}.
\newblock


\bibitem[\protect\citeauthoryear{{\c{C}}olak, Lima, and
  Gonz{\'a}lez}{{\c{C}}olak et~al\mbox{.}}{2016}]%
        {2020-sub-31}
\bibfield{author}{\bibinfo{person}{Serdar {\c{C}}olak},
  \bibinfo{person}{Antonio Lima}, {and} \bibinfo{person}{Marta~C
  Gonz{\'a}lez}.} \bibinfo{year}{2016}\natexlab{}.
\newblock \showarticletitle{Understanding congested travel in urban areas}.
\newblock \bibinfo{journal}{\emph{Nature}} \bibinfo{volume}{7},
  \bibinfo{number}{1} (\bibinfo{year}{2016}), \bibinfo{pages}{1--8}.
\newblock


\bibitem[\protect\citeauthoryear{Dean}{Dean}{1999}]%
        {2020-sub-10}
\bibfield{author}{\bibinfo{person}{Brian~C Dean}.}
  \bibinfo{year}{1999}\natexlab{}.
\newblock \emph{\bibinfo{title}{Continuous-time dynamics shortest path
  algorithms}}.
\newblock \bibinfo{thesistype}{Ph.D. Dissertation}.
  \bibinfo{school}{Massachusetts Institute of Technology}.
\newblock


\bibitem[\protect\citeauthoryear{Dean}{Dean}{2004}]%
        {2020-sub-08}
\bibfield{author}{\bibinfo{person}{Brian~C Dean}.}
  \bibinfo{year}{2004}\natexlab{}.
\newblock \showarticletitle{Shortest paths in FIFO time-dependent networks:
  Theory and algorithms}.
\newblock \bibinfo{journal}{\emph{Rapport Technique}} (\bibinfo{year}{2004}),
  \bibinfo{pages}{13}.
\newblock


\bibitem[\protect\citeauthoryear{Demiryurek, Banaei-Kashani, and
  Shahabi}{Demiryurek et~al\mbox{.}}{2010}]%
        {m24}
\bibfield{author}{\bibinfo{person}{Ugur Demiryurek}, \bibinfo{person}{Farnoush
  Banaei-Kashani}, {and} \bibinfo{person}{Cyrus Shahabi}.}
  \bibinfo{year}{2010}\natexlab{}.
\newblock \showarticletitle{A case for time-dependent shortest path computation
  in spatial networks}. In \bibinfo{booktitle}{\emph{ACM SIGSPATIAL}}.
  \bibinfo{pages}{474--477}.
\newblock


\bibitem[\protect\citeauthoryear{{Department for Transport}}{{Department for
  Transport}}{[n.d.]}]%
        {highwaycode_headway}
\bibfield{author}{\bibinfo{person}{{Department for Transport}}.}
  \bibinfo{year}{[n.d.]}\natexlab{}.
\newblock \bibinfo{title}{The Highway Code}.
\newblock
\newblock
\urldef\tempurl%
\url{https://www.gov.uk/guidance/the-highway-code/}
\showURL{%
\tempurl}


\bibitem[\protect\citeauthoryear{Dijkstra et~al\mbox{.}}{Dijkstra
  et~al\mbox{.}}{1959}]%
        {dijkstra1959note}
\bibfield{author}{\bibinfo{person}{Edsger~W Dijkstra} {et~al\mbox{.}}}
  \bibinfo{year}{1959}\natexlab{}.
\newblock \showarticletitle{A note on two problems in connexion with graphs}.
\newblock \bibinfo{journal}{\emph{Numerische mathematik}} \bibinfo{volume}{1},
  \bibinfo{number}{1} (\bibinfo{year}{1959}), \bibinfo{pages}{269--271}.
\newblock


\bibitem[\protect\citeauthoryear{El~Hatri and Boumhidi}{El~Hatri and
  Boumhidi}{2017}]%
        {el2017traffic}
\bibfield{author}{\bibinfo{person}{Chaimae El~Hatri} {and}
  \bibinfo{person}{Jaouad Boumhidi}.} \bibinfo{year}{2017}\natexlab{}.
\newblock \showarticletitle{Traffic management model for vehicle re-routing and
  traffic light control based on Multi-Objective Particle Swarm Optimization}.
\newblock \bibinfo{journal}{\emph{Intelligent Decision Technologies}}
  \bibinfo{volume}{11}, \bibinfo{number}{2} (\bibinfo{year}{2017}),
  \bibinfo{pages}{199--208}.
\newblock


\bibitem[\protect\citeauthoryear{Fairclough, May, and Carter}{Fairclough
  et~al\mbox{.}}{1997}]%
        {fairclough1997effect}
\bibfield{author}{\bibinfo{person}{Stephen~H Fairclough},
  \bibinfo{person}{Andrew~J May}, {and} \bibinfo{person}{C Carter}.}
  \bibinfo{year}{1997}\natexlab{}.
\newblock \showarticletitle{The effect of time headway feedback on following
  behaviour}.
\newblock \bibinfo{journal}{\emph{Accident Analysis \& Prevention}}
  \bibinfo{volume}{29}, \bibinfo{number}{3} (\bibinfo{year}{1997}),
  \bibinfo{pages}{387--397}.
\newblock


\bibitem[\protect\citeauthoryear{Geolink}{Geolink}{2015}]%
        {Porto2015}
\bibfield{author}{\bibinfo{person}{Geolink}.} \bibinfo{year}{2015}\natexlab{}.
\newblock \bibinfo{booktitle}{\emph{ECML/PKDD 15: Taxi Trajectory Prediction}}.
\newblock
\urldef\tempurl%
\url{http://www.geolink.pt/ecmlpkdd2015-challenge/dataset.html}
\showURL{%
Retrieved August 15, 2019 from \tempurl}


\bibitem[\protect\citeauthoryear{George and Shekhar}{George and
  Shekhar}{2008}]%
        {m2}
\bibfield{author}{\bibinfo{person}{Betsy George} {and} \bibinfo{person}{Shashi
  Shekhar}.} \bibinfo{year}{2008}\natexlab{}.
\newblock \bibinfo{booktitle}{\emph{Time-Aggregated Graphs for Modeling
  Spatio-Temporal Networks}}.
\newblock \bibinfo{pages}{191--212}.
\newblock


\bibitem[\protect\citeauthoryear{Guo, Li, Zhang, Ding, Curtmola, and
  Borcea}{Guo et~al\mbox{.}}{2020}]%
        {2020-sub-28}
\bibfield{author}{\bibinfo{person}{Chang Guo}, \bibinfo{person}{Demin Li},
  \bibinfo{person}{Guanglin Zhang}, \bibinfo{person}{Xiaoning Ding},
  \bibinfo{person}{Reza Curtmola}, {and} \bibinfo{person}{Cristian Borcea}.}
  \bibinfo{year}{2020}\natexlab{}.
\newblock \showarticletitle{Dynamic Interior Point Method for Vehicular Traffic
  Optimization}.
\newblock \bibinfo{journal}{\emph{IEEE Transactions on Vehicular Technology}}
  \bibinfo{volume}{69}, \bibinfo{number}{5} (\bibinfo{year}{2020}),
  \bibinfo{pages}{4855--4868}.
\newblock


\bibitem[\protect\citeauthoryear{Hart, Nilsson, and Raphael}{Hart
  et~al\mbox{.}}{1968}]%
        {hart1968formal}
\bibfield{author}{\bibinfo{person}{Peter~E Hart}, \bibinfo{person}{Nils~J
  Nilsson}, {and} \bibinfo{person}{Bertram Raphael}.}
  \bibinfo{year}{1968}\natexlab{}.
\newblock \showarticletitle{A formal basis for the heuristic determination of
  minimum cost paths}.
\newblock \bibinfo{journal}{\emph{IEEE Transactions on Systems Science and
  Cybernetics}} \bibinfo{volume}{4}, \bibinfo{number}{2}
  (\bibinfo{year}{1968}), \bibinfo{pages}{100--107}.
\newblock


\bibitem[\protect\citeauthoryear{Jahn, M{\"o}hring, Schulz, and
  Stier-Moses}{Jahn et~al\mbox{.}}{2005}]%
        {jahn2005system}
\bibfield{author}{\bibinfo{person}{Olaf Jahn}, \bibinfo{person}{Rolf~H
  M{\"o}hring}, \bibinfo{person}{Andreas~S Schulz}, {and}
  \bibinfo{person}{Nicol{\'a}s~E Stier-Moses}.}
  \bibinfo{year}{2005}\natexlab{}.
\newblock \showarticletitle{System-optimal routing of traffic flows with user
  constraints in networks with congestion}.
\newblock \bibinfo{journal}{\emph{Operations Research}} \bibinfo{volume}{53},
  \bibinfo{number}{4} (\bibinfo{year}{2005}), \bibinfo{pages}{600--616}.
\newblock


\bibitem[\protect\citeauthoryear{Jeong, Jeong, Lee, Oh, and Du}{Jeong
  et~al\mbox{.}}{2015}]%
        {jeong2015saint}
\bibfield{author}{\bibinfo{person}{Jaehoon Jeong}, \bibinfo{person}{Hohyeon
  Jeong}, \bibinfo{person}{Eunseok Lee}, \bibinfo{person}{Tae Oh}, {and}
  \bibinfo{person}{David~HC Du}.} \bibinfo{year}{2015}\natexlab{}.
\newblock \showarticletitle{SAINT: Self-adaptive interactive navigation tool
  for cloud-based vehicular traffic optimization}.
\newblock \bibinfo{journal}{\emph{IEEE Transactions on Vehicular Technology}}
  \bibinfo{volume}{65}, \bibinfo{number}{6} (\bibinfo{year}{2015}),
  \bibinfo{pages}{4053--4067}.
\newblock


\bibitem[\protect\citeauthoryear{Levinson}{Levinson}{2018}]%
        {levinson_2018}
\bibfield{author}{\bibinfo{person}{David Levinson}.}
  \bibinfo{year}{2018}\natexlab{}.
\newblock \bibinfo{title}{How much time is spent at traffic signals?}
\newblock
\newblock
\urldef\tempurl%
\url{https://transportist.org/2018/03/06/how-much-time-is-spent-at-traffic-signals/}
\showURL{%
\tempurl}


\bibitem[\protect\citeauthoryear{Madkour, Aref, Rehman, Rahman, and
  Basalamah}{Madkour et~al\mbox{.}}{2017}]%
        {madkour2017survey}
\bibfield{author}{\bibinfo{person}{Amgad Madkour}, \bibinfo{person}{Walid~G
  Aref}, \bibinfo{person}{Faizan~Ur Rehman}, \bibinfo{person}{Mohamed~Abdur
  Rahman}, {and} \bibinfo{person}{Saleh Basalamah}.}
  \bibinfo{year}{2017}\natexlab{}.
\newblock \showarticletitle{A survey of shortest-path algorithms}.
\newblock  (\bibinfo{year}{2017}).
\newblock
\showeprint[arxiv]{1705.02044}


\bibitem[\protect\citeauthoryear{May}{May}{1990}]%
        {may1990traffic}
\bibfield{author}{\bibinfo{person}{Adolf~Darlington May}.}
  \bibinfo{year}{1990}\natexlab{}.
\newblock \bibinfo{booktitle}{\emph{Traffic flow fundamentals}}.
\newblock \bibinfo{publisher}{Prentice Hall}, Chapter~10.
\newblock


\bibitem[\protect\citeauthoryear{Motallebi, Xie, Tanin, and
  Ramamohanarao}{Motallebi et~al\mbox{.}}{2020}]%
        {motallebi2020streaming}
\bibfield{author}{\bibinfo{person}{Sadegh Motallebi}, \bibinfo{person}{Hairuo
  Xie}, \bibinfo{person}{Egemen Tanin}, {and} \bibinfo{person}{Kotagiri
  Ramamohanarao}.} \bibinfo{year}{2020}\natexlab{}.
\newblock \showarticletitle{Streaming route assignment with prior temporal
  traffic data}. In \bibinfo{booktitle}{\emph{ACM SIGSPATIAL IWCTS}}.
  \bibinfo{pages}{1--10}.
\newblock


\bibitem[\protect\citeauthoryear{Nguyen, Karunasekera, Kulik, Tanin, Zhang,
  Zhang, Xie, and Ramamohanarao}{Nguyen et~al\mbox{.}}{2015}]%
        {nguyen2015randomized}
\bibfield{author}{\bibinfo{person}{Uyen~TV Nguyen}, \bibinfo{person}{Shanika
  Karunasekera}, \bibinfo{person}{Lars Kulik}, \bibinfo{person}{Egemen Tanin},
  \bibinfo{person}{Rui Zhang}, \bibinfo{person}{Haolan Zhang},
  \bibinfo{person}{Hairuo Xie}, {and} \bibinfo{person}{Kotagiri
  Ramamohanarao}.} \bibinfo{year}{2015}\natexlab{}.
\newblock \showarticletitle{A randomized path routing algorithm for
  decentralized route allocation in transportation networks}. In
  \bibinfo{booktitle}{\emph{ACM SIGSPATIAL IWCTS}}. \bibinfo{pages}{15--20}.
\newblock


\bibitem[\protect\citeauthoryear{Pan, Khan, Popa, Zeitouni, and Borcea}{Pan
  et~al\mbox{.}}{2012}]%
        {pan2012proactive}
\bibfield{author}{\bibinfo{person}{Juan Pan}, \bibinfo{person}{Mohammad~A
  Khan}, \bibinfo{person}{Iulian~Sandu Popa}, \bibinfo{person}{Karine
  Zeitouni}, {and} \bibinfo{person}{Cristian Borcea}.}
  \bibinfo{year}{2012}\natexlab{}.
\newblock \showarticletitle{Proactive vehicle re-routing strategies for
  congestion avoidance}. In \bibinfo{booktitle}{\emph{IEEE Conference on
  Distributed Computing in Sensor Systems}}. \bibinfo{pages}{265--272}.
\newblock


\bibitem[\protect\citeauthoryear{Sperb}{Sperb}{2010}]%
        {2020-sub-11}
\bibfield{author}{\bibinfo{person}{R~Campi Sperb}.}
  \bibinfo{year}{2010}\natexlab{}.
\newblock \bibinfo{booktitle}{\emph{Solving time-dependent shortest path
  problems in a database context}}.
\newblock \bibinfo{publisher}{University of Twente}.
\newblock


\bibitem[\protect\citeauthoryear{TLC}{TLC}{2015}]%
        {NYC2020}
\bibfield{author}{\bibinfo{person}{New York~City TLC}.}
  \bibinfo{year}{2015}\natexlab{}.
\newblock \bibinfo{booktitle}{\emph{TLC Trip Record Data}}.
\newblock
\urldef\tempurl%
\url{https://www1.nyc.gov/site/tlc/about/tlc-trip-record-data.page}
\showURL{%
Retrieved January 15, 2019 from \tempurl}


\bibitem[\protect\citeauthoryear{Wang, Li, Chen, and Ni}{Wang
  et~al\mbox{.}}{2009}]%
        {wang2009speed}
\bibfield{author}{\bibinfo{person}{Haizhong Wang}, \bibinfo{person}{Jia Li},
  \bibinfo{person}{Qian-Yong Chen}, {and} \bibinfo{person}{Daiheng Ni}.}
  \bibinfo{year}{2009}\natexlab{}.
\newblock \showarticletitle{Speed-Density Relationship: From Deterministic to
  Stochastic}. In \bibinfo{booktitle}{\emph{TRB Annual Meeting}}.
\newblock


\bibitem[\protect\citeauthoryear{Wang, Li, and Tang}{Wang
  et~al\mbox{.}}{2019}]%
        {2020-sub-09}
\bibfield{author}{\bibinfo{person}{Yong Wang}, \bibinfo{person}{Guoliang Li},
  {and} \bibinfo{person}{Nan Tang}.} \bibinfo{year}{2019}\natexlab{}.
\newblock \showarticletitle{Querying shortest paths on time dependent road
  networks}.
\newblock \bibinfo{journal}{\emph{PVLDB}} \bibinfo{volume}{12},
  \bibinfo{number}{11} (\bibinfo{year}{2019}), \bibinfo{pages}{1249--1261}.
\newblock


\bibitem[\protect\citeauthoryear{Wardrop}{Wardrop}{1952}]%
        {wardrop1952road}
\bibfield{author}{\bibinfo{person}{John~Glen Wardrop}.}
  \bibinfo{year}{1952}\natexlab{}.
\newblock \showarticletitle{Some theoretical aspects of road traffic research}.
\newblock \bibinfo{journal}{\emph{Proceedings of the Institution of Civil
  Engineers}} \bibinfo{volume}{1}, \bibinfo{number}{3} (\bibinfo{year}{1952}),
  \bibinfo{pages}{325--362}.
\newblock


\bibitem[\protect\citeauthoryear{Yen}{Yen}{1971}]%
        {m17}
\bibfield{author}{\bibinfo{person}{Jin~Y Yen}.}
  \bibinfo{year}{1971}\natexlab{}.
\newblock \showarticletitle{Finding the k shortest loopless paths in a
  network}.
\newblock \bibinfo{journal}{\emph{Management Science}} \bibinfo{volume}{17},
  \bibinfo{number}{11} (\bibinfo{year}{1971}), \bibinfo{pages}{712--716}.
\newblock


\bibitem[\protect\citeauthoryear{Zhan and Noon}{Zhan and Noon}{1998}]%
        {zhan1998shortest}
\bibfield{author}{\bibinfo{person}{F~Benjamin Zhan} {and}
  \bibinfo{person}{Charles~E Noon}.} \bibinfo{year}{1998}\natexlab{}.
\newblock \showarticletitle{Shortest path algorithms: an evaluation using real
  road networks}.
\newblock \bibinfo{journal}{\emph{Transportation Science}}
  \bibinfo{volume}{32}, \bibinfo{number}{1} (\bibinfo{year}{1998}),
  \bibinfo{pages}{65--73}.
\newblock


\bibitem[\protect\citeauthoryear{Zhang, Jiang, and Ma}{Zhang
  et~al\mbox{.}}{2012}]%
        {zhang2012improved}
\bibfield{author}{\bibinfo{person}{Wei Zhang}, \bibinfo{person}{Chong Jiang},
  {and} \bibinfo{person}{Yunxiang Ma}.} \bibinfo{year}{2012}\natexlab{}.
\newblock \showarticletitle{An improved Dijkstra algorithm based on pairing
  heap}. In \bibinfo{booktitle}{\emph{International Symposium on Computational
  Intelligence and Design}}, Vol.~\bibinfo{volume}{2}.
  \bibinfo{pages}{419--422}.
\newblock


\end{thebibliography}

\end{document}